\documentclass[mathleft
]{an}
\usepackage{graphicx}
\usepackage{times}
\usepackage{epsfig}
\usepackage{amsmath}
\usepackage{graphicx}
\usepackage{amssymb}
\begin{document}

\Pagespan{666}{671}

\Yearpublication{2014}%
\Yearsubmission{2014}%
\Month{}%
\Volume{335}%
\Issue{6/7}%

\title{Effects of strong magnetic fields on the population of hyperon stars}

\author{R.O. Gomes\inst{1,2}\fnmsep\thanks{Corresponding author:
  \email{rosana.gomes@ufrgs.br}\newline}  \and V. Dexheimer\inst{3}
  \and C.A.Z. Vasconcellos\inst{1,4}
   \newline}

\titlerunning{Effects of strong magnetic fields on the population of hyperon stars}
\authorrunning{R.O. Gomes et al.}

\institute{Instituto de F\'isica, Universidade Federal do Rio
Grande do Sul, Av. Bento Gon\c{c}alves 9500, CEP  91501-970, Porto
Alegre, RS, Brazil \and Frankfurt Institute for Advances Studies,
Ruth-Moufang-Strasse 1 60438 Frankfurt am Main \and Department of
Physics, Kent State University, Kent OH, 44242 USA \and ICRANet,
P.zza della Repubblica 10, 65122 Pescara, Italy }

\received{2014 Apr 30} \accepted{2014 May 30} \publonline{2014 Aug 01}


\abstract{In this contribution we study the effects of strong
magnetic fields on the particle population of neutron stars
with hyperon degrees of freedom in their composition. The star
matter is described by a multi-component model with parameterized
baryon-meson interaction couplings. We study the magnetic effects
on the equation of state (EoS) due to the Landau quantization,
assuming a density dependent static magnetic field that reaches
about $10^{19}\,G$ in the center of the star. The
Tolman-Oppenheimer-Volkoff equations are solved in order to
understand the dependence of the mass-radius relation and hyperon
population on the magnetic field intensity assuming
different interaction coupling schemes.}

\maketitle

\section{Introduction}

 In recent years, magnetars have attracted increasing attention due
to their extremely powerful magnetic fields and the emission of
high-energy electromagnetic radiation, particularly X-rays and
gamma rays,
 broadening our conceptions about the formation
 and evolution of neutron stars.
  Magnetars are neutron
 stars with strong magnetic fields, whose intensities are larger
 than the critical value at which the ener\-gy between Landau
 quantized levels of electrons equals their rest mass.

The critical value for the intensity of the magnetic field in
neutron stars can be estimated by assuming  that electronic Landau
levels, i.e., the discrete energy values  $E_{n}$ of
 the orbits of electrons under the action of an external magnetic field $B$, may be described by the
 quantum states of the
three dimensional harmonic oscillator
$E_n = \hslash \omega \left(n + \frac{3}{2} \right) ,$
with $n = 0, 1, 2, 3, ..$. From this expression, we obtain the
separation $\Delta E_{n, n+1}$ between successive Landau le\-vels,
labelled respectively as $n$ and $n+1$:
\begin{equation}
\Delta E_{n, n+1} = \hslash \omega  \, .
\end{equation}
Taking the classical cyclotron frequency $\omega$ of the orbital
motion of an electron in a magnetic field
$ \omega = \frac{eB}{m_ec} \, $, we obtain $B = m_e  c
\frac{\omega}{e} \, ,$
 where $m_e$ represents the electron mass and $e$ its electrical
 charge, and assuming $\Delta E_{n, n+1} = \hslash
\omega = m_e c^2$ allows one to estimate the critical value $B_C$
of the magnetic field
\begin{equation}\label{eq:Bquantum}
B_C = \frac{m_e^2 c^3}{\hslash e} \sim 4.4 \times 10^{13}~{\rm G} \, .
\end{equation}

Compact objects with surface magnetic fields as strong as
$10^{14}-10^{15}$ G, have been observed and this has drawn the
attention to the study of the equation of state (EoS) of hadronic
matter in the presence of strong magnetic fields. The intensity of
magnetic fields in the interior region of magnetars are expected
to be even stronger, maybe reaching $10^{19}$ G (Ferrer et al
2010).

According to Rea \& Esposito (2011), the large magnetic field of a
magnetar causes extreme activity such as large outbursts (during
which their X-ray flux output can increase by several orders of
magnitude) accompanied by short ener\-getic X-ray bursts. However,
there is some controversy about the origin of such intense
magnetic fields inside neutron stars. Candidates for magnetars are
often referred as Soft Gamma Repeaters (SGRs) or Anomalous X-Ray
Pulsars (AXPs), depending on the characteristics of their
electromagnetic emissions. For a review of this topic see Woods \&
Thompson (2006) and Mereghetti (2008). There are however
alternative models to the magnetar scenario, for instance the
white dwarf model which describes the SGRs and AXPs powered by the
rotational energy of a massive, fast rotating, highly magnetized
white dwarf with surface fields in the range $10^8$--$10^{10}$~G
(see Malheiro et al 2012, for details).

Strong magnetic fields may also be very important for neutron star
formation. Stars more massive than about $8$ solar masses and less
massive than about $25$ solar masses in the beginning of their
lives are supposed to be the prime candidates for the standard
formation scenario behind core-collapse supernovae. During this
process, when the star core reaches nuclear densities, nuclear
forces and the neutron degeneracy pressure sharply increase the
core internal pressure, causing the core to bounce sending a shock
wave throughout the star (Bethe \& Brown 1998). The nature of the
bounce mechanism depends upon the equation of state of the core:
accordingly, a stiffer equation of state causes the star to bounce
more quickly but also more weakly than a softer equation of state.
As an alternative scenario, magnetic fields may deposit enough
energy on the protoneutron star and drive the explosion.
Within the current uncertainties in these mechanisms, strong
magnetic fields produce massive neutron stars, in agreement
with the as\-tro\-no\-mi\-cal observations.

Finally,  the presence of stable matter at super-nuclear densities
in neutron stars makes then essential tools for the understanding of
hadronic physics. An important feature in neutron
stars is the possibility of new hadronic degrees of freedom,
in addition to neutrons and protons. Hyperons, i.e.,
baryons with strange content, are hypothesized to appear in the
dense interior of neutron stars at about twice the nuclear
saturation density. This can easily be explained by the Pauli
principle, since as the baryon density increases, the same occurs
with the Fermi momentum and the Fermi energy, thereby enlarging
the phase space of the baryon sector. Note that the predominant effect of
the presence of hyperons is to soften the equation of state of
nuclear matter, due to the decay of energetic nucleons at the top
of the Fermi sea into hyperons at lower Fermi levels,
ge\-ne\-ra\-ting as a
consequence neutron stars with smaller masses.
Nevertheless, is not a sufficient argument to ignore the hyperons,
as discussed in detail in Ref. (Buballa et al. 2014).

In this work, we study the combination of the two aforementioned ingredients in
order to understand the hyperon role in the composition of stellar
matter and the respective mass-radius relationship, and compare
our predictions with a recent observation that accounts for a
neutron star mass of the order of two solar masses (Demorest, P.
et al, ~2010). We examine the role of the presence of hyperons in
the core of neutron stars with emphasis on effects that can be
attributed to the general multi-species composition of nuclear
matter. With this goal, we study the effects of intense magnetic
fields on the population and also in global properties of neutron
stars, using a relativistic effective model with parameterized
couplings involving the interaction between nucleons, hyperons and
mesons.

\section{Model description} \label{Model}

We consider in this paper a scalar version of the effective
re\-la\-ti\-vistic QHD-model with parameterized couplings.
This model represents an extended compilation of other QHD models found in
the literature (Vasconcellos ~2012). The complete
expression of our scalar QHD interaction Lagrangian
exhausts the whole fundamental baryon octet ($n$, $p$, $\Sigma^-$,
$\Sigma^0$, $\Sigma^+$, $\Lambda$, $\Xi^-$, $\Xi^0$) but considers
Yukawa-type interaction terms only involving the $\omega$ and the
$\varrho$ meson fields. Additionally,  our approach includes
non-Yukawa many-body interaction terms through nonlinear
self-couplings involving the scalar-isoscalar $\sigma$ meson.
For more complete description see Ref. (Vasconcellos et al. ~2014).


In our approach, the self-interaction terms are simulated by means
of a parametrization of the coupling constants of the theory. For
certain va\-lues of the parameters, the treatment
adopted in this work reproduces the same predictions for global
properties of neutron stars as most of the models based on
Yukawa-type couplings in\-vol\-ving the $\sigma$, $\omega$ and
$\varrho$ mesons. For other choices of the parameters, our
approach allows the description of the effects of density self-correlations of
higher orders involving the scalar-isoscalar meson $\sigma$
on global properties of neutron stars.

In the mean field approximation, the parameterized coupling Lagrangian density with scalar
self-energy insertions, $\Sigma^{s \ast}_{B \lambda \, 0}$, is defined as (Vasconcellos 2012)
\begin{eqnarray}
\!\!\!\!\!\!\!\! {\cal L}_{\lambda} & \! = \! &
\frac{1}{2}m_\sigma^2\sigma_0^2  +  \frac{1}{2}m_\omega^2\omega_0^2
 +   \frac{1}{2} m_\rho^2 \varrho_{03}^2
\\ & \! +  \! &  \sum_l \bar\psi_l \,(i \gamma_\mu
\partial^\mu +q_e \gamma_\mu A^\mu  -  m_l)\,\psi_l \nonumber
\\  & \! + \! & \sum_B \bar{\psi}_B (i \gamma_\mu \partial^\mu \! + \! q_e \gamma_\mu A^\mu
  -   g_{\omega B} \gamma^0 \omega_0) \psi_B \nonumber \\  &  \! - \! &
 \sum_B  \!
 \bar{\psi}_B  ( \frac{1}{2} g_{\varrho B}
 \gamma^0 \!
\tau^{(3)} \! \varrho_{03}   +    M_{B \lambda} \Sigma^{s \ast}_{B \lambda \, 0} ) \psi_B
 \, , \nonumber \label{Lagrangian}
\end{eqnarray}
where the subscripts $B$ and $l$ represent, respectively, the
baryon octet ($n$, $p$, $\Lambda^0$, $\Sigma^-$, $\Sigma^0$,
$\Sigma^+$, $\Xi^-$, $\Xi^0$) and lepton ($e^-$, $\mu^-$) species.
The electromagnetic interaction is introduced by the $A^\mu$
field, with the coupling intensity given by the particles
electrical charge $q_e$. In this expression, $g_{\Phi}$, with
$\Phi =\sigma, \omega, \varrho$, re\-pre\-sents the effective
baryon-meson coupling constants and $\sigma_0$, $\omega_0$ and
$\varrho_{03}$ are the mean field values of the $\sigma$, $\omega$
and $\mbox{\boldmath$\varrho$}_{\mu}$ meson fields. The
self-energy insertion $ \Sigma^{s \ast}_{B \lambda \, 0}$ is
defined as
\begin{eqnarray}  \Sigma^{s \ast}_{B \lambda \, 0} & = &  m^*_{B \lambda}
  \equiv  \left( 1 + \frac{g_{\sigma B} \sigma_0}{\lambda M_B} \right)^{-\lambda}  \,
  , \label{SE}
\end{eqnarray}
and gives rise to the effective parameterized baryon mass
\begin{eqnarray}
\!\!\!\!\!\!\! M^*_{B \lambda} & = & M_B m^*_{B \lambda}
\\
 & \simeq & \! M_B \! \left( \!\! 1 -
\frac{g_{\sigma B}\sigma_0}{M_B} +  \left( \!
\begin{array}{c} \lambda \\ 2  \end{array} \right) \left( \! \frac{g_{\sigma B} \sigma_0}{\lambda M_B}  \right)^2   \right) \! + \! {\cal O}(3) \,
,  \nonumber \\ \label{M} \nonumber
\end{eqnarray}
for $ \frac{g_{\sigma B} \sigma_0}{\lambda M_B} <<1 $; in this
expression $ \left(\begin{array}{c} \lambda \\ 2  \end{array}
\right) $ represents the generalized binomial coefficients of the
expansion; we also emphasize in this expression the direct
dependence of the effective baryon mass on the $\lambda$ parameter
of the model. Finally, note that our method is equivalent to replace the original Yukawa scalar
coupling $g_{\sigma B} \sigma$ by the parameterized scalar coupling
\begin{equation}
g_{\sigma B} \sigma  \to  g^*_{\sigma B \lambda} \sigma_0 \equiv
m^*_{B \lambda} g_{\sigma B} \sigma_0 \, .
\end{equation}
Properties of the fields considered in our formulation are
presented in table (\ref{campos}).

In a previous contribution to this volume (Vasconcellos et al.
2014), we have determined the equation of state,  population
profiles and the mass-radius relation for families of neutron
stars with hyperon content. In this contribution, we
additionally consider the effects of strong magnetic fields
on the properties of neutron stars with hyperon degrees of freedom.

\begin{center}
\begin{table}[t]
 \caption{Properties of the fields considered in the formulation.
In what follows, we use the abbreviations: ISS: isoscalar-scalar;
ISV: isoscalar-vector; IVV:
isovector-vector.}
\begin{center}
\begin{tabular}{llclc}
\hline
Fields & Classification & Particles & Coupling   & Mass \\
 & & & Constants & (MeV) \\
 \hline
$\psi_B$ & Baryons & N, \,$\Lambda$,& N/A & 939, 1116, \\
         &           &              $\Sigma$, \,$\Xi$   &     &       1193, 1318\\
$\psi_l$ & Leptons & $e^-$, $\mu^-$& N/A &0,5,\,106 \\
$\sigma $ & ISS-meson  & $\sigma $ & $g^*_{\sigma_B}$ & 550 \\
$\omega_\mu $ & ISV-meson  & $\omega $ &$g^*_{\omega_B}$ &782 \\
$\mbox{\boldmath$\varrho$}_{\mu}$ & IVV-meson  & $\rho$ & $g^*_{\varrho_B}$&770 \\
\hline\hline
\end{tabular}
\label{campos}
\end{center}
\end{table}
\end{center}

\section{Coupling constants}

When constraining the coupling constants of our approach, we use the
conventional procedure, which considers that the predictions of
an effective theory must be in line with two-body scattering results and bulk static
properties data of nuclear matter at saturation density. This means
that the theory should reproduce the main properties of finite nuclei.
After this, the next step comprises extrapolating the theory to the
high density regime.

We constrain in the following the $\lambda$ parameter in order to
describe the effective mass of the nucleon and compression modulus
at saturation density. We emphasize that our model predicts the values of
nuclear saturation properties in good agreement with literature
(shown in Table (\ref{CC})). We have adopted for the saturation density of nuclear
matter $\rho_0= 0.17$ fm$^{-3}$ and for the binding energy
$\epsilon_B = -16.0$ MeV. The $\lambda$ parameter is constrained
to describe the nucleon effective mass at saturation between
$[0.70-0.78]$ MeV. The isovector coupling constant $g_{\varrho N}$
on the other hand was chosen to describe the symmetry energy
coefficient $a_{sym}=32.5$ MeV (for more details see Haensel,
Potekhin \& Yakovlev ~2007). Based on this, we introduce a coherent
set of nuclear-meson coupling constants $g_{\sigma N}$, $g_{\omega N}$ for symmetric matter in the saturation density range.

In the high density regime, as already mentioned, hyperon degrees of freedom must be
taken into account. However, since these fields are not present in
nuclear matter at saturation density, it is not possible to
determine the hyperon-meson (HM) couplings directly. In most
models found in the literature, hyperon degrees of freedom appear
in nuclear matter at around two times the nuclear saturation
density $\rho_0$, as for instance, in relativistic mean-field
models (Wang et al. ~2006),
in non-relativistic potential models (Dabrowski \& Rozynek
 ~2010),
in the quark-meson coupling model (Whittenbury et
al. ~2013),
in relativistic Hartree-Fock models (Huber et al. ~1998),
in Brueckner-Hartree-Fock calculations (Shternin, Baldo \&
Haensel ~2013),
and within chiral effective Lagrangian (Banik  et al. ~2004).
Nevertheless, the details of the hyperon composition of neutron
star matter are rather sensitive to the chosen hyperon potentials.

\begin{table}[h,t,b]
  \caption{\label{cc} Coupling parameters of our approach.}
\begin{tabular}{cccccc} \small
$\lambda$ & $(\frac{M^*}{M})_{_0}$ &  $K_{_0}$ (MeV) & $(\frac{g_{\sigma
B}}{m_{\sigma}})^2$ & $(\frac{g_{\omega B}}{m_{\omega}})^2$ &
$(\frac{g_{\varrho B}}{m_{\varrho}})^2$ \\
  \hline \hline
    & &  & & & \\
  0.06 & 0.70  & 262   &  11.87  & 6.49  & 3.69  \\
  0.10 & 0.75  & 226   &  10.42  & 5.10  & 3.94  \\
  0.14 & 0.78  & 216   &  9.51   & 4.32   & 4.07 \\
\end{tabular}
\label{CC}
\end{table}
As the density of the system increases, it becomes  more energetically favorable for the system to create hyperon
species than increase the energy levels of the nucleon. In order to des\-cri\-be the hyperon couplings, we define
the hyperon-meson couplings as $g_{\eta B} = \chi_{\eta B}\,g_{\eta N}$ for $\eta = \sigma,\omega,\varrho$.
In this work, in order to constrain the hyperon population, we use the three hyperon coupling models that follows:
\begin{itemize}
\item {\bf HYS(1)} (Moszkowski 1974): it is based on the different nature of hyperons with respect to nucleons:
\begin{equation}
\chi_{\sigma B}=\chi_{\omega B}=\chi_{\varrho B}=\sqrt{2/3};
\end{equation}
\item \textbf{HYS(2)} (Pal et al. 1999): it is based on the counting quark rule.
\begin{equation}
\chi_{\sigma\Lambda}=2/3,\quad\chi_{\sigma\Sigma}=2/3,\quad\chi_{\sigma\Xi}=1/3,
\end{equation}
\begin{equation}
\chi_{\omega\Lambda}=2/3,\quad\chi_{\omega\Sigma}=2/3,\quad\chi_{\omega\Xi}=1/3,
\end{equation}
\begin{equation}
\chi_{\varrho\Lambda}=0,\quad\chi_{\varrho\Sigma}=2,\quad\chi_{\varrho\Xi}=1;
\end{equation}
\item \textbf{HYS(3)} (Rufa et al. 1990; Glendenning \& Moszkowski
1991): it is based on experimental analysis of
$\Lambda$-hypernucleus data
\begin{equation}
\chi_{\sigma B}=\chi_{\sigma\Lambda} \, ; \, \chi_{\omega B} = \chi_{\omega\Lambda} \, ; \, \chi_{\varrho B}=0 \, ,
\end{equation}
where the binding energy is given by:
\begin{eqnarray}
(B/A)_{\Lambda} & =  & \chi_{\omega B}\,(g_{\omega N}\,\omega_{0}) + \chi_{\sigma B} \, (m_{\Lambda}^{*} -m_{\Lambda})
\nonumber \\
& = & -28 MeV \, ,
\end{eqnarray}
and
\begin{equation}
\chi_{\sigma \Lambda} = 0.75 \, .
\end{equation}
\end{itemize}

\begin{figure}[h]
\centering \epsfig{file=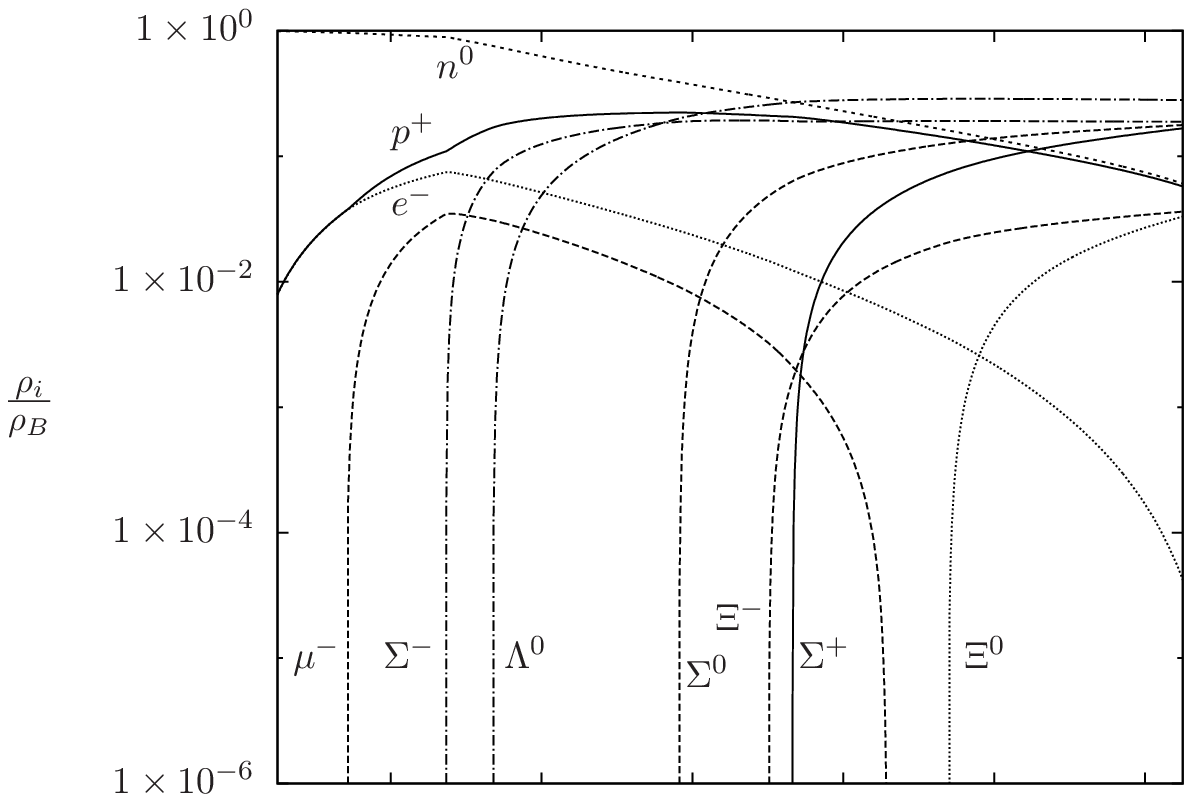,height=2.3in}
\hspace{0.3cm} \centering
\epsfig{file=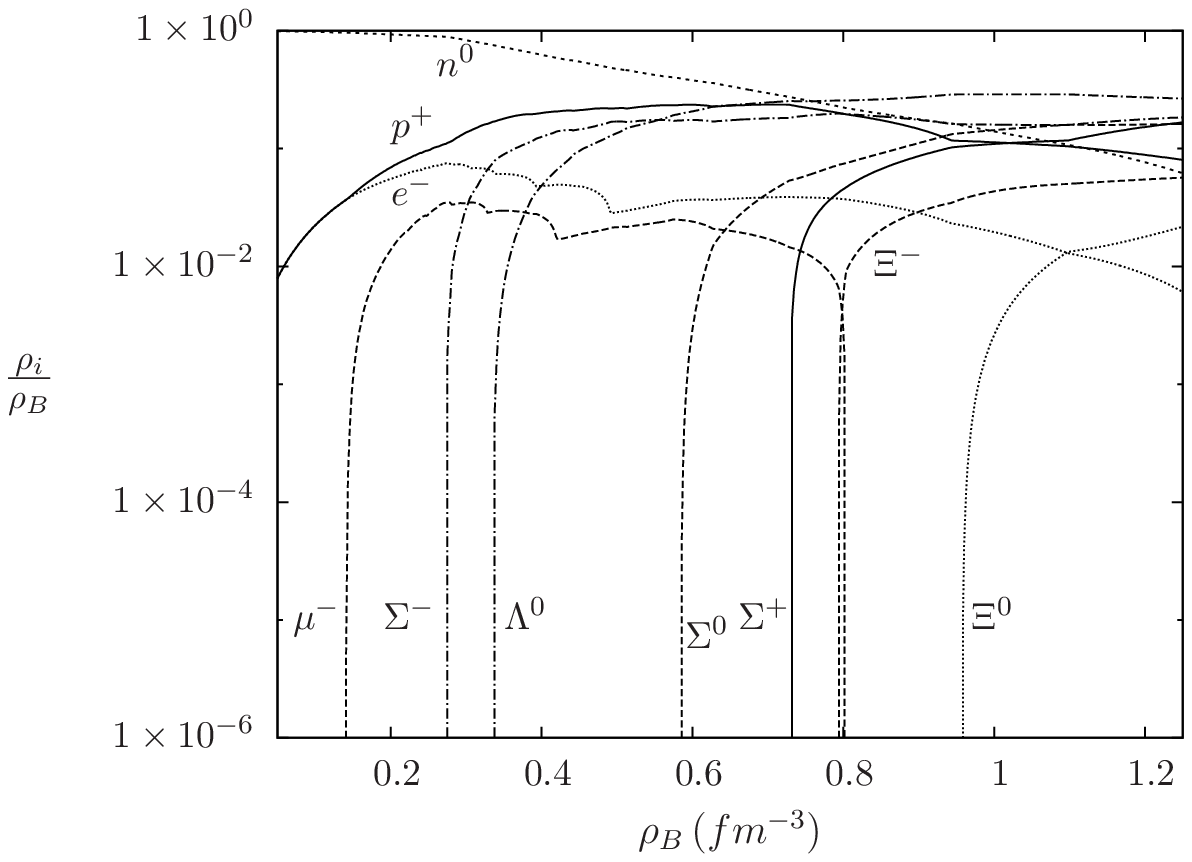,height=2.3in}
\caption{\small{Particles densities (divided by baryon density) as
a function of the baryon density, using the hyperon coupling model
$HYS(1)$ and the parametrization from table \ref{cc}
($\lambda=0.06$). The top panel  does not have magnetic field
contributions. The bottom panel includes magnetic field
contributions, with $B_c=10^{19}\,G$.}} \label{1}
\end{figure}
\begin{figure}[h]
\centering \epsfig{file=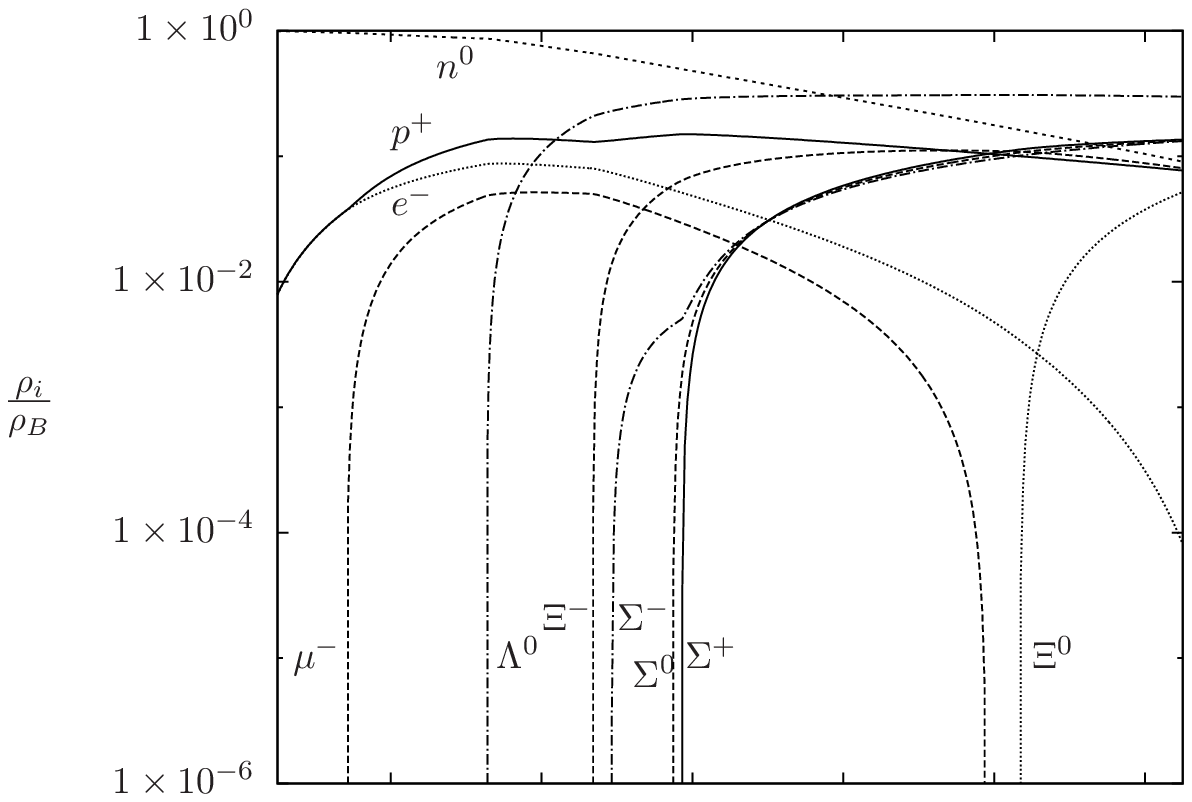,height=2.3in}
\hspace{0.3cm} \centering
\epsfig{file=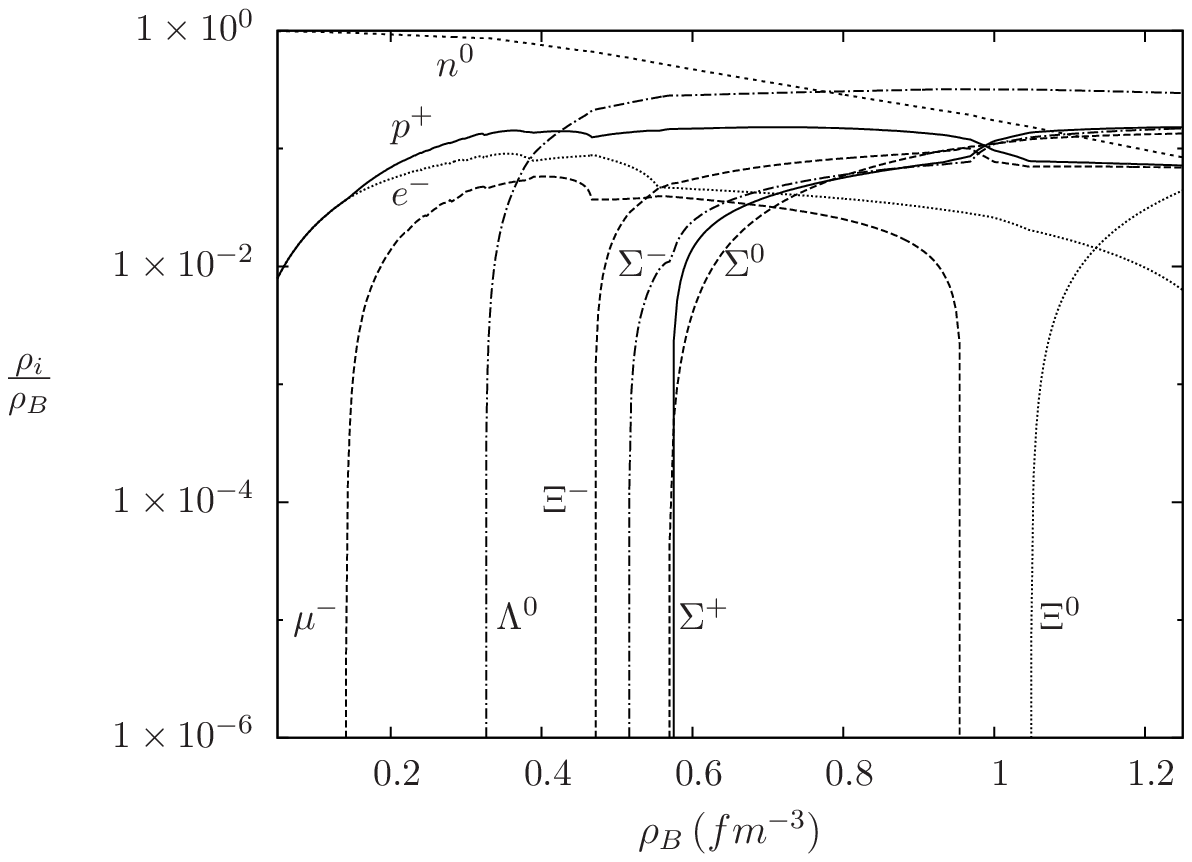,height=2.3in} \caption{\small{Same
as figure \ref{1}, but for the hyperon coupling model HYS(2).}}
\label{2}
\end{figure}
\begin{figure}[h]
\centering \epsfig{file=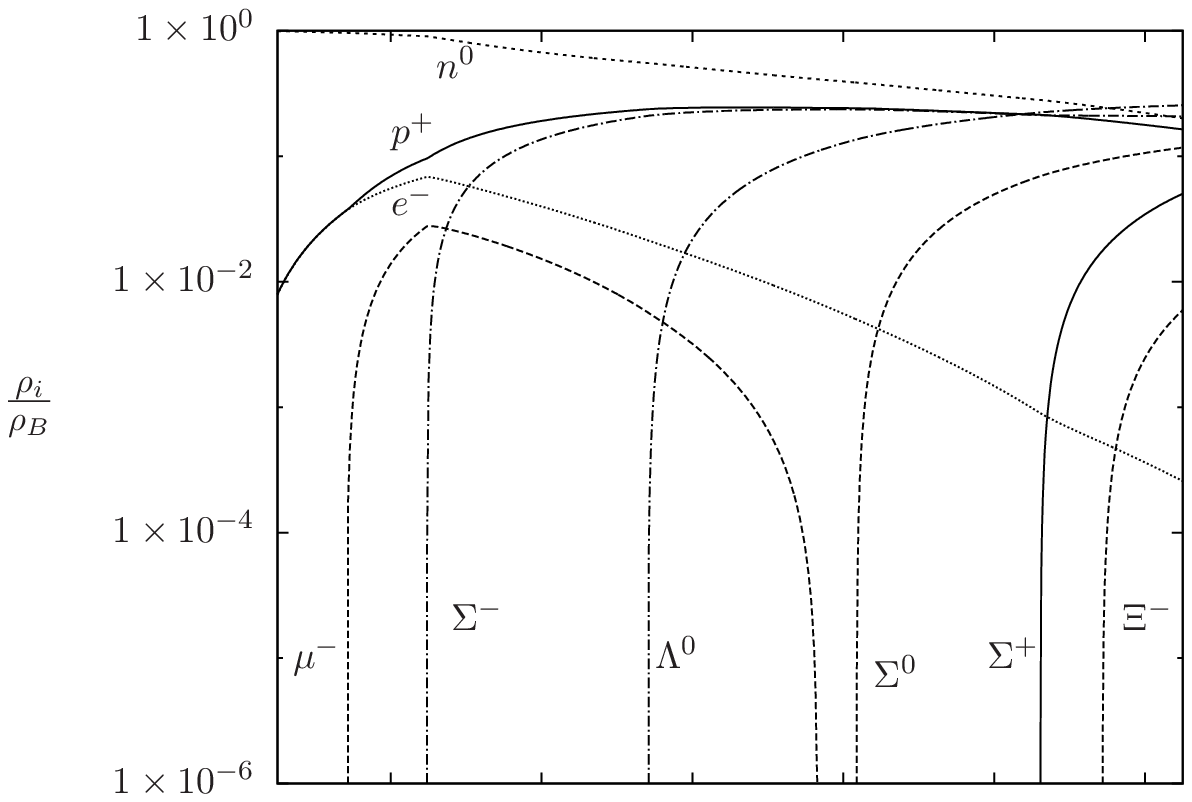,height=2.3in}
\hspace{0.3cm} \centering
\epsfig{file=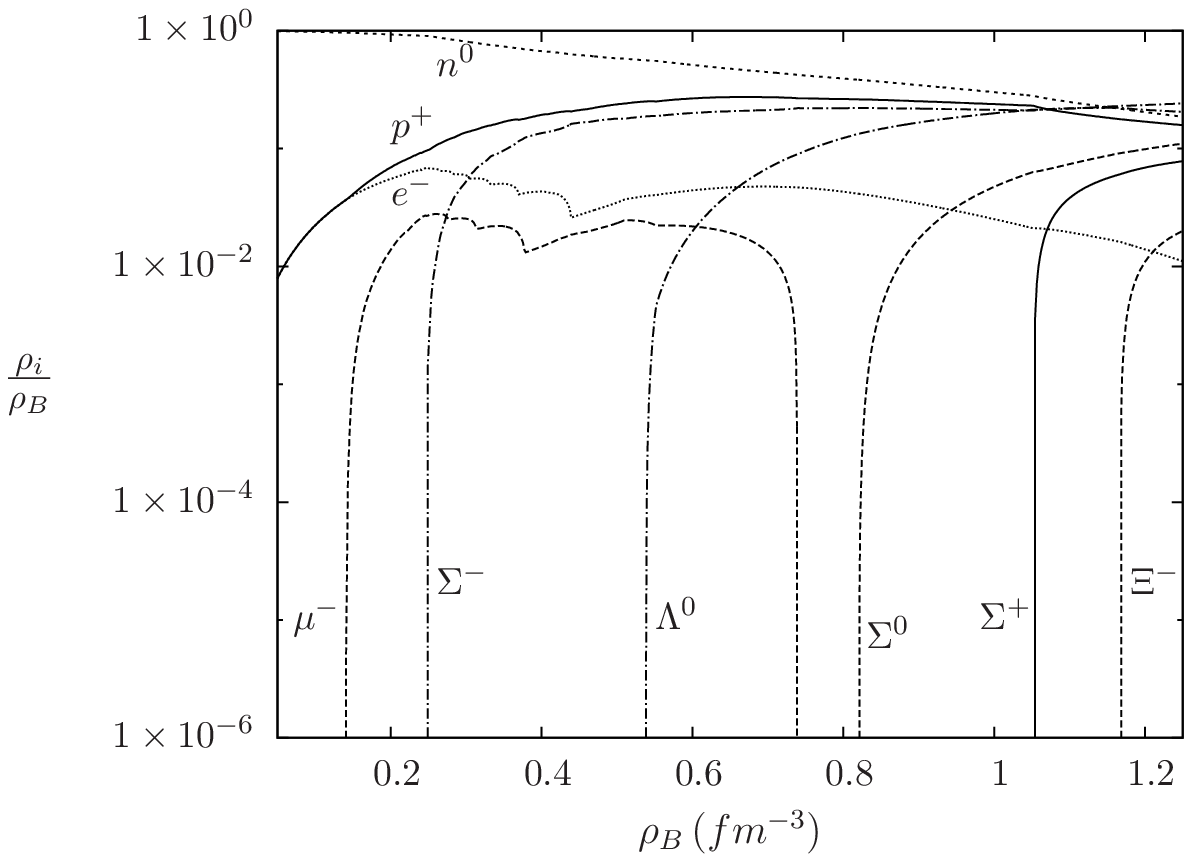,height=2.3in} \caption{\small{Same
as figure \ref{1}, but for the hyperon coupling model HYS(3).}}
\label{3}
\end{figure}
\begin{figure}[h]
\centering \epsfig{file=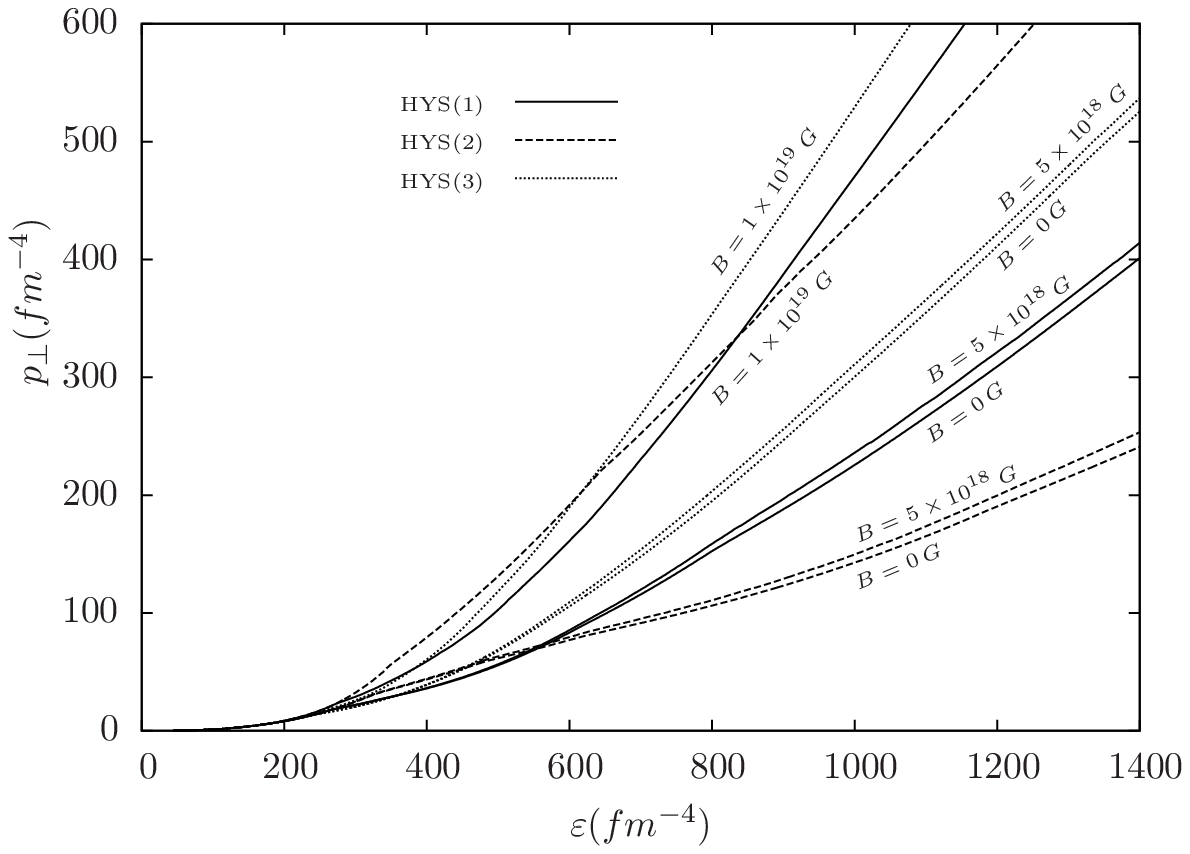,height=2.3in} \hspace{0.3cm}
\centering \epsfig{file=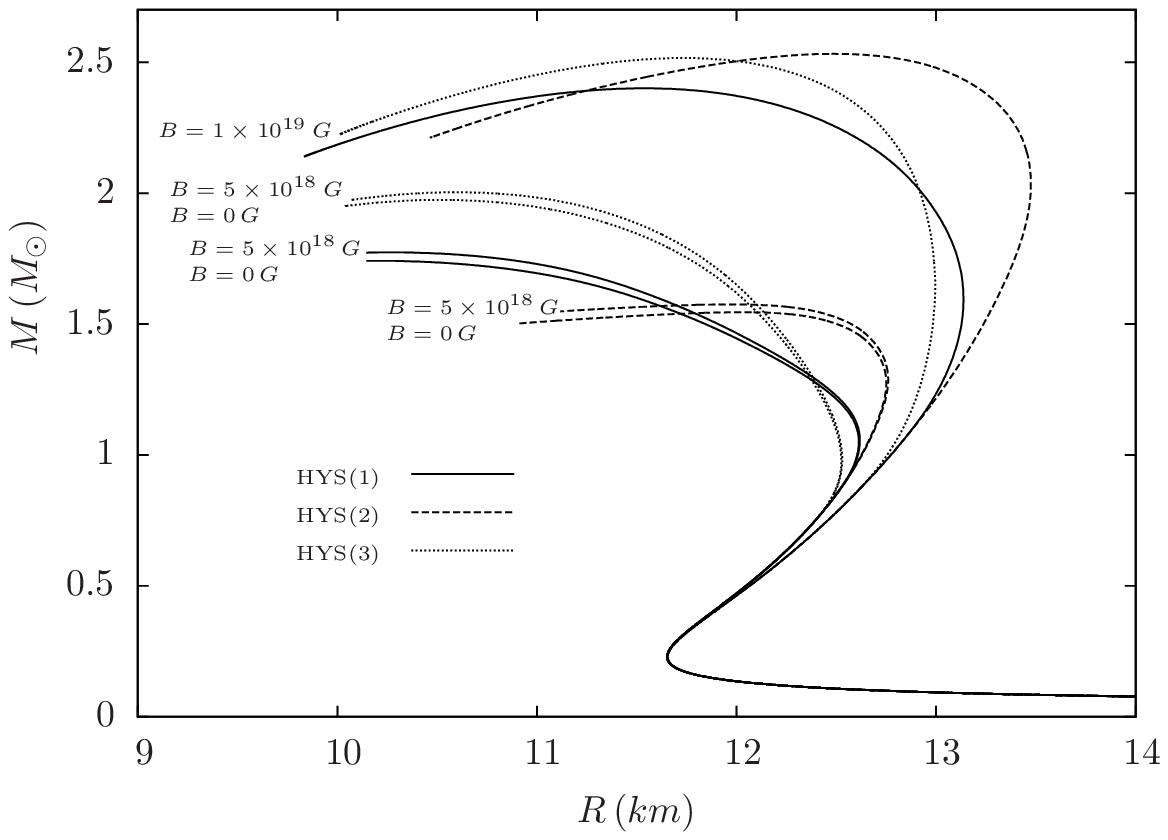,height=2.3in} \caption{EoS (top
panel) and mass-radius diagram (bottom panel) for different
central magnetic fields $B_c$, and different hyperon coupling
model. All curves use the first parametrization on table \ref{cc}
($\lambda=0.06$).} \label{eos_tov}
\end{figure}
\begin{figure}[h]
\centering \epsfig{file=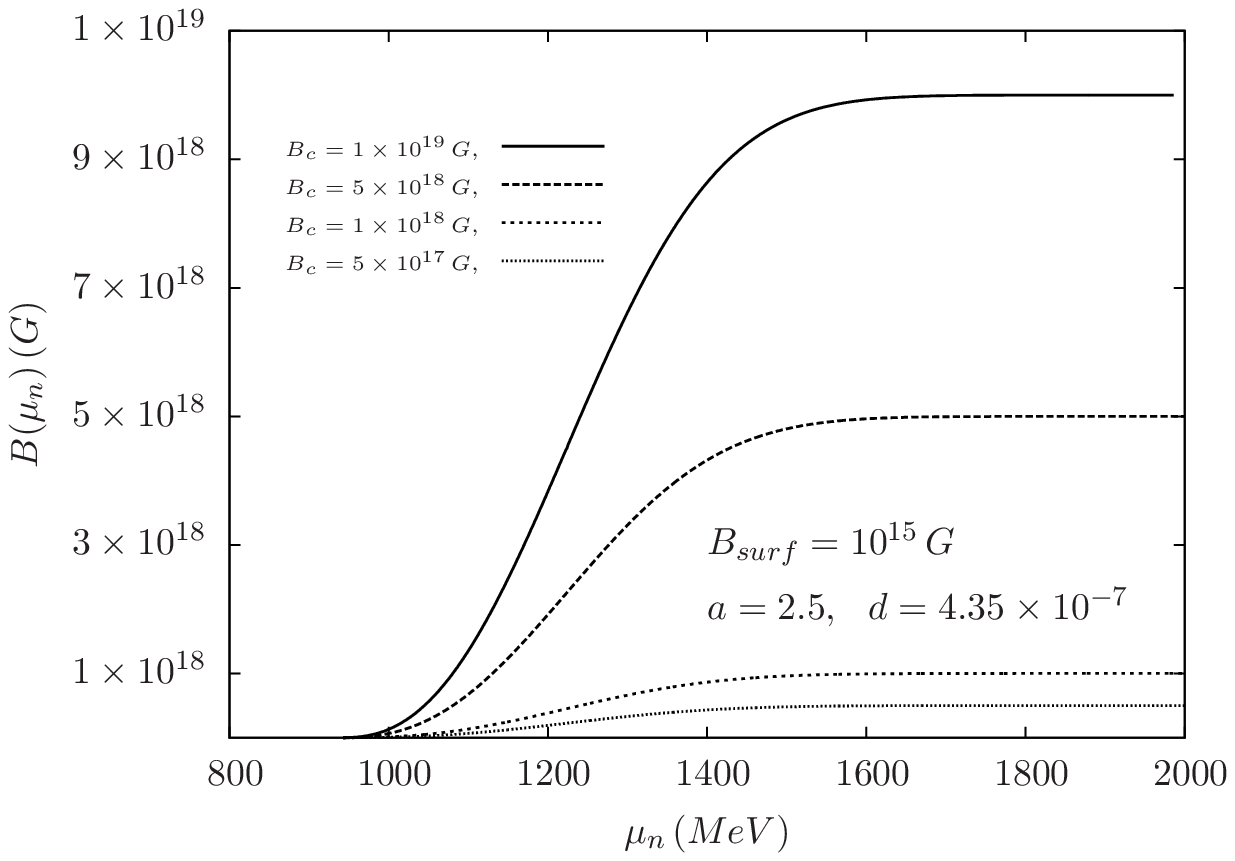,height=2.3in} \caption{Magnetic
field dependence on baryon chemical potential. We use the
parameterization: $a=2.5$, $b=4.35\times10^{-7}$, and a surface
magnetic field of $B_{surf}=10^{15}\,G$, the highest one measured
(Dexheimer 2011).}
\end{figure}

\section{Magnetic field and Landau quantization}

In the presence of an external magnetic field $B$ in the $z$ direction, the orbital motion of charged particles (baryons and leptons)
is described by the Landau quantization energy spectra
\begin{equation}
E_{\nu, i}=\sqrt{k_{z_i}^{2}+ m^{2}_{i} +2\nu\vert q_{e}\vert{B}},
\end{equation}
with $i = B, l$. Note that in the expression above
we must consider the effective mass of baryons.
The Landau levels are double degenerated except for the fundamental state, and are enumerated by $\nu$.
We calculate the EoS from the spatial and temporal contributions of the energy-momentum tensor of the system, following the solutions
already calculated in the literature by  (Broderick, Prakash \& Lattimer 2002).

The equation of state  $P_{mag} = P_{mag}(\varepsilon_{mag})$  may be expressed as
\begin{eqnarray}
\!\!\!\!\!\!\!\!\! \varepsilon_{mag} & \! = \! &
 \frac{1}{2} m_{\sigma}^2\sigma_0^2 + \frac{1}{2}m_{\omega}^2\omega_0^2 + \frac{1}{2}m_{\varrho}^2\varrho^2_{03}  \\ & \! + \! &
\frac{\vert q_{e}\vert{B}}{4\pi^2}  \! \sum_{b, \nu} \eta(\nu) \! \left[ \! k_{F_{b},\nu}\,\mu_{b}^* \! + \!
\overline{m}_{b,\nu}^{\, 2}\,ln  \left( \! \frac{\mu_{b}^* \! + \! k_{F_{b},\nu}}{\overline{m}_{b,\nu}} \! \right) \! \right] \nonumber \\ & \! + \! &
\frac{\vert q_{e}\vert{B}}{4\pi^2} \!  \sum_{l, \nu} \eta(\nu) \! \left[ \! k_{F_{l},\nu}\,\mu_{l}   +
\overline{m}_{l,\nu}^{\, 2} \, ln \left( \! \frac{\mu_{l} \! + \! k_{F_{l},\nu}}{\overline{m}_{l,\nu}} \! \right) \! \right], \nonumber
\end{eqnarray}
with
\begin{eqnarray}
\!\!\!\!\!\!\!\!\! P_{mag} & \! = \! & - \frac{1}{2}m_{\sigma}^2\sigma_0^2+ \frac{1}{2}m_{\omega}^2\omega_0^2+\frac{1}{2}m_{\varrho}^2\varrho^2_{03} \\ & + &
\frac{\vert q_{e}\vert{B}}{4\pi^2}  \! \sum_{b, \nu} \eta(\nu) \! \left[ \! k_{F_{b},\nu}\,\mu_{b}^* \! - \!
\overline{m}_{b,\nu}^{\, 2} \,ln \! \left(\! \frac{\mu_{b}^* \! + \! k_{F_{b},\nu}}{\overline{m}_{b,\nu}} \! \right) \! \right] \nonumber \\ & + &
\frac{\vert q_{e}\vert{B}}{4\pi^2}  \sum_{l, \nu} \eta(\nu) \! \left[ k_{F_{l},\nu}\,\mu_{l} -
\overline{m}_{l,\nu}^2\, ln \left( \! \frac{\mu_{l} \! + \! k_{F_{l},\nu}}{\overline{m_{l,\nu}}} \! \right) \! \right].  \nonumber
\end{eqnarray}

We also allow the pressure to become anisotropic in the presence
of a  strong magnetic field, as pointed out by Refs.
(Perez~Martinez, Perez~Rojas \& Mosquera~Cuesta 2008; Strickland,
Dexheimer \& Menezes, 2012; Paulucci et al 2011)
\begin{eqnarray}
 \varepsilon & = &  \sum_{B,l} \varepsilon_{mag} + \frac{B^2}{2} \, \, ;
 P_{\Vert} = \sum_{b,l} P_{mag} - \frac{B^2}{2} \, , \nonumber \\
 P_{\bot} & = &  \sum_{b,l} P_{mag} + \frac{B^2}{2} - B\mathcal{M} \, ,
\end{eqnarray}
where the magnetization is calculated as $\mathcal{M}= \partial P_{mag}/\partial B $.

We assume a magnetic field with chemical potential dependence (Bandyopadhyay et al.
1997; Dexheimer et al. 2012):
\begin{equation}\label{eq:Bfunc}
B(\mu)= B_{surf}+B_{c}\left[1-\exp\left(-b\,(\mu_n-938)^{a}\right)\right],
\end{equation}
where $B_{c}$ represents the magnetic field in the high $\mu_n$ limit and
 the parameters $a$ and $b$ determine how fast the magnetic field increases towards the center of the star.

\section{Results and conclusions}

We model particle populations considering the conditions of beta-equilibrium, charge neutrality and baryon number conservation.
Also, we assume that the baryon chemical potential
suffers a shift by the vector mesons due to the nuclear interaction:
\begin{eqnarray}
\mu^*_i = q_{b_{i}}\,\mu_n - q_{e_{i}}\,\mu_e -g_{\omega}\omega - g_{\varrho}\varrho\frac{\bf \tau}{2}.
\end{eqnarray}

In addition, our hadronic model considers magnetic effects on
the hyperon population of neutron stars with strong magnetic fields.
As can be seen in the top panel of Figs. (\ref{1}), (\ref{2}) and (\ref{3}), the different hyperon coupling models
reproduce quite different star populations.

The chemical potential of particles is responsible for their
appearance in the system and it is calculated based on the
coupling constants with the mesons. More precisely, the coupling
with vector mesons lowers the chemical potential, allowing the
baryons to be created at lower densities, among other conditions.
In particular, the HYS(3) proposes that all hyperon species have a
zero coupling constant with the meson $\varrho$, implying a higher
chemical potential. The consequence of this choice of model is the
appearance of hyperon species only at high densities - in
comparison with HYS(1) and HYS(2) - which have the direct effect
of stiffening the EoS, as shown in the top panel of Fig.
(\ref{eos_tov}).

Moreover, the introduction of magnetic fields in the model raises
several new issues concerning the star population which are discussed in the
following. The first one is the change in the order that the
hyperons appear as a function of density, like for example the
$\Sigma^+$ and $\Xi^{-}$ in the bottom panel of Fig. (\ref{1}).
The second one is the change in the amount of charged particle
species, like for example the electrons at high densities in the
bottom panel of Fig. (\ref{3}). The third issue is the wriggles
that appear due to the Landau quantization and integer values of
$\nu$, which can be clearly seen for example in the bottom panel
of Fig. (\ref{2}). These would be even more different (from the
case in which the magnetic field was ignored) if we had included
the effects of anomalous magnetic moment in the model, as done in
Refs. (Dexheimer,  Negreiros \& Schramm~2012; Isayev \& Yang~2013
). In such case, the population of particles with different spins
becomes quite different, and in same cases, fully polarized.

The difference in particle population caused by the magnetic field
also affects the EoS, as seen in the top panel of Fig.
(\ref{eos_tov}). Besides the effect of clearly stiffening all EoS
at high densities, the magnetic field exchanges the order
(according to stiffness) of the EoS of different hyperon coupling
models at intermediate densities. For example, the curve for the
HYS(2) with $B_c=10^{19}$~G at $\epsilon \sim 600$~fm$^{-4}$,
which was the softer EOS without magnetic field but becomes the
stiffer one in the presence of strong magnetic field. This in
turn, makes the HYS(2)-EoS to reproduce more massive stars
(compared with the other hyperon coupling models) but only in the
presence of strong magnetic fields. This can be seen in the bottom
panel of Fig. (\ref{eos_tov}).

All these effects are up to some extent model dependent.
Nevertheless, they still contain relevant physical conclusions,
such as the importance of better understanding the magnetic field
dependence with density and on the structure of neutron stars. The
dependence of the EoS and particles population on the different
hyperon-meson couplings models at high densities also highlights
the fact that a better analysis of the hyperon-meson coupling needs
to be explored in order to quantify the effect of these
uncertainties on the observable properties of neutron stars.

\end{document}